\documentclass[twocolumn,showpacs,preprintnumbers,amsmath,amssymb]{revtex4}
\usepackage{graphicx}
\usepackage{dcolumn}
\usepackage{bm}
\begin{document}
\newcommand{\etal}{et al. } 

\title{Focusing of laser-generated ion beams by a plasma cylinder:
similarity theory and the thick lens formula}

\author{S.~Gordienko$^{1,2}$, T.~Baeva$^1$, A.~Pukhov$^1$} 

\affiliation{$^1$Institut f\"ur Theoretische Physik I,
Heinrich-Heine-Universit{\"a}t D\"usseldorf, D-40225, Germany \\
$^2$L.~D.~Landau Institute for Theoretical Physics, Moscow, Russia
}

\date{\today}

\begin{abstract}
\noindent 

It is shown that plasma-based optics can be used to guide and focus highly
divergent laser-generated ion beams. A hollow cylinder is considered,
which initially contains a hot electron population. Plasma streaming
toward the cylinder axis maintains a focusing electrostatic field due
to the positive radial pressure gradient. The cylinder works as thick
lens, whose parameters  are obtained from similarity
theory for freely expanding plasma in cylindrical
geometry. Because the lens parameters are energy dependent, the
lens focuses a selected energy range of ions and works as a
monochromator. Because the focusing is due to the quasineutral part of
the expanding plasma, the lens parameters depend 
on the hot electron temperature $T_e$ only, and not their density. 

\end{abstract}


\maketitle

\section{Introduction}

Laser-driven ion sources
\cite{LLNL,Roth,focusing,Vulcan,Vulcan1,RothFI,ions3d} are 
considered to be the 
%%%%%%%%%%%%%%%%%%%%%%%%%%%%%%%%%%
hot candidates for various important applications
in nuclear physics, medicine, biology, material sciences, plasma field
tomography \cite{Umstadter,Ledingham,Romagnani}.
When multi-terawatt laser pulses are 
%%%%%%%%%%%%%%%%%%%%%%%%%%%%%%%%%%%%
shot on solid state targets, 
copious amounts of multi-MeV ions - both protons and highly charged
heavier ions - are generated \cite{Hegelich}. These laser-generated
ion beams have picosecond durations and originate from a few
micrometer wide virtual source. 
However, the
laser-generated ions are highly divergent and usually are emitted
within a cone with some 10-30 degrees opening angle. In addition,
they have broad energy spectra. These facts may impede numerous
applications for the laser-generated ion beams unless appropriate
optics and monochromatizing systems are developed.

Because of their high divergence, one needs very
strong fields to collimate the ion beams. Such fields exist
only in plasma. However, one cannot exploit the standard technique of
self-induced magnetic plasma lensing that is widely
used to focus conventionally accelerated ion beams. The reason is that
the laser-produced ion beams are charge neutral, i.e. they contain 
electrons that compensate the ion charge and current.

\section{Similarity theory of expanding plasma}

\begin{figure}
\centerline{\includegraphics[width=8cm,clip]{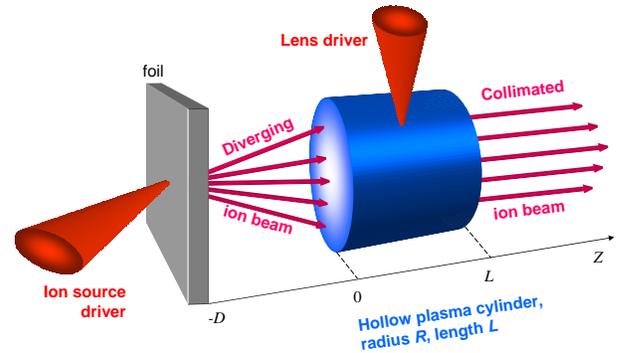}}
\caption{(color online). Geometry of the ion optics element for
focusing of laser-generated ion beams. One laser beam generates an
ion beam from the rear side of the irradiated foil. Another laser beam
hits a hollow cylinder, where a hot electron population is generated.
The cylinder has the radius $R$, the length $L$ and is located at the
distance $D$ from the foil. The radial electric field of the plasma 
inside the cylinder collimates the ion beam.} \label{fig:lens} 
\end{figure}

In the present work we consider ion beam focusing by plasma which
already contains a quasistatic electric field. 
The experimental configuration is the following, Fig.~\ref{fig:lens},
\cite{LaserPhysHHU}: a laser produced ion beam originates at
the point $z=-D$, $x=y=0$ and propagates toward a
hollow metal cylinder of the radius $R$ and length $L$, $L>R$. The
axis of the cylinder coincides with the $z$-axis. At the same time, 
the second laser pulse is shot at the
cylinder. This second pulse generates a population of hot electrons,
which penetrate through the metal and spread very fast 
over the inner surface of the cylinder. They exit into vacuum and
generate a cloud of space charge at the inner surface. The electric
field of this space charge is large enough to ionize the material and
to create plasma. 
As a result at the initial
moment we have a cylindrical plasma 
layer with high electron 
temperature $T_e$ and low ion temperature $T_i\approx 0$.
The plasma begins to expand toward the cylinder
axis due to the TNSA (target normal sheath acceleration) mechanism
\cite{WilksIons}.
Normally, the cylinder surface is
covered by a thin layer of hydrogen-rich substances. Being the lightest
ions, protons are accelerated first and the plasma is usually an
electron-proton one regardless of the particular chemical mixture
of the cylinder itself.

%%%%%%%%%%%%%%%%%%%%%%%%%%%%%%%%
Plasma dynamics is described by the couple of Vlasov's equations for
electrons and ions and the Maxwell equations

\begin{eqnarray}
\label{Vlasov}
\frac{\partial f_e(t,{\bf p}, {\bf r})}{\partial t}& +& {\bf
v}\cdot\frac{\partial f_e(t,{\bf p},
{\bf r})}{\partial {\bf r}}-{e{\bf E}}\cdot\frac{\partial f_e(t,{\bf p},
{\bf r})}{\partial {\bf p}}=0\nonumber\\
\frac{\partial f_i(t,{\bf p}, {\bf r})}{\partial t} & + & {\bf
v}\cdot\frac{\partial f_i(t,{\bf p},
{\bf r})}{\partial {\bf r}}+{e\mathcal{Z}_i{\bf E}}\cdot\frac{\partial f_i(t,{\bf p},
{\bf r})}{\partial {\bf p}}=0, \nonumber \\
\nabla\cdot {\bf E}&=&4\pi e\left( \mathcal{Z}_i\int f_i\,d^3{\bf
p}-\int f_e\,d^3{\bf p}\right), 
\end{eqnarray}

\noindent where $f_e$ and $f_i$ are the electron and ion distribution functions
respectively. We do not include magnetic field in (\ref{Vlasov}) since
the cylindrical symmetry of the expansion prohibits magnetic fields
generation. 

The initial conditions for Eqs. (\ref{Vlasov}) are

\begin{eqnarray}
\label{InitialConditions}
f_e(t=0,{\bf p},{\bf r})&=&f_0(c{\bf p}/T_e,{\bf r}/R,{\bf r}/d),\nonumber\\
f_i(t=0,{\bf p},{\bf r})&=&F_0({\bf r}/R,{\bf r}/d)\delta({\bf p}),
\end{eqnarray}

\noindent where $f_0$ and $F_0$ are initial distributions of electrons and ions,
$d$ and $R$ are cylinder thickness and radius respectively, $R>d$. For
the initial distributions one reads

\begin{eqnarray}
\label{InitialConditionsDensity}
\int f_0(c{\bf p}/T_e,{\bf r}/R,{\bf r}/d)\,d{\bf p}d{\bf R}&=&2\pi R d L n_e,\nonumber\\
\int F_0({\bf r}/R,{\bf r}/d)\delta({\bf p})d{\bf p}d{\bf R}&=&2\pi R d L n_e/\mathcal{Z}_i,
\end{eqnarray}

\noindent where the cylinder length $L\gg R$, $n_e$ is the average
electron density. The multiplier 
$1/\mathcal{Z}_i$ is due to the condition of plasma charge neutrality. 

In the following we consider the case of relativistic electron
temperatures and non-relativistic ions: $M_ic^2\gg T_e\gg
m_ec^2$. Thus, we exploit the ultrarelativistic approximation ${\bf
v}=c{\bf p}/|{\bf p}|$ for the electron velocities, while
for the ions we get  ${\bf v}={\bf p}/M_i$. Here $M_i$ and $m_e$ are
the ion and electron masses respectively. 

We introduce new dimensionless variables 

\begin{equation}
\label{DimensionlessVariables}
{\hat {\bf r}}={\bf r}/R,~~{\hat t}=t/\tau,~~\tau=R/c_s,~~{\hat {\bf
E}}=eR{\bf E}/{T_e},
\end{equation}

\noindent where $c_s=\sqrt{T_e/M_i}$ is the ion sound velocity. Since the electrons
are ultra-relativistic and ions are non-relativistic, we are forced to introduce
different dimensionless normalization for electron and ion components:

\begin{eqnarray}
\label{DimensionlessElectron}
{\hat {\bf p_e}}=c{\bf p_e}/T_e,~&~{\hat f}_e=\left(T_e/c\right)^3f_e/n_e,\\
\label{DimensionlessIons}
{\hat{\bf p_i}}={\bf p_i}/\sqrt{M_i T_e},~&~{\hat
f}_i=\left({M_iT_e}\right)^{3/2}f_i/n_e.
\end{eqnarray}

\noindent  We rewrite the Vlasov equations (\ref{Vlasov}) in these variables: 

\begin{eqnarray}
\label{VlasovDimensionless}
\alpha_c\frac{\partial {\hat f}_e({\hat t},{\hat {\bf p_e}}, {\hat
{\bf r}})}{\partial {\hat t}} + \frac{{\hat{ \bf 
p_e}}}{|{\hat {\bf p_e}}|}\frac{\partial {\hat f}_e({\hat t},{\hat
{\bf p_e}},  
{\hat {\bf r}})}{\partial {\hat {\bf r}}}-{\hat {\bf
E}}\frac{\partial {\hat f}_e({\hat t},{\hat {\bf p_e}}, 
{\hat {\bf r}})}{\partial {\hat{\bf p_e}}}&=&0\nonumber\\
\frac{\partial \hat{f}_i({\hat t},{\hat {\bf p_i}},{\hat {\bf
r}})}{\partial \hat{t}} + {\hat{\bf p_i}}\frac{\partial {\hat
f}_i({\hat t},{\hat{\bf p_i}}, {\hat{\bf r}})}{\partial {\hat{\bf
r}}}+{\mathcal{Z}_i{\hat{\bf E}}}\frac{\partial {\hat f}_i({\hat
t},{\hat {\bf p_i}}, {\hat {\bf r}})}{\partial {\bf p_i}}&=&0,\nonumber\\ 
\alpha_D\nabla\cdot\hat{\bf E}=4\pi e\left(\mathcal{Z}_i\int{\hat
f}_i\,d^3{\hat{\bf p}}-\int {\hat f}_e\,d^3{\hat{\bf p}}\right),  & &
\end{eqnarray}

\noindent The normalized Vlasov-Maxwell equations
(\ref{VlasovDimensionless}) reveal that the the system dynamics depend
on five dimensionless parameters. The first parameter is the ion charge
$\mathcal{Z}_i$. The next two parameters are the normalized sound
speed $\alpha_c = c_s/c$ and the normalized Debye length
$\alpha_D=\lambda_D^2/4\pi R^2$, where $\lambda_D^2=4\pi
T_e/e^2n_e$. These two parameters define plasma dynamic properties. 
The remaining two parameters $d/R$ and $L/R$ come from the
initial system geometry. 
We are interested in the cylindrical geometry and drop out the
parameter $L/R\rightarrow +\infty$. 

%A non-trivial result related to
%Eqs. (\ref{VlasovDimensionless}) is that 
%the dimensionless parameter $N_D=n_e\lambda_D^3$ allowed by the dimensional reasonings does not
%enter the theory and does not affect the plasma dynamics. One can readily check that this parameter
%is functionally independent from $\alpha_c$ and $\alpha_D$. 

Thus, the parametric dependencies can be written as:

\begin{eqnarray}
\label{DimensionlessFunctions}
f_e&=&\frac{n_ec^3}{T_e^3}{\hat f}_e\left(\frac{t}{\tau},\frac{\bf r}{R},\frac{c{\bf p}}{T_e}, \frac{d}{R},\mathcal{Z}_i, \alpha_c,\alpha_D\right),\\
f_i&=&\frac{n_e}{(MT_e)^{3/2}}{\hat f}_i\left(\frac{t}{\tau},\frac{\bf
r}{R},\frac{{\bf p}}{(MT_e)^{1/2}}, \frac{d}{R},\mathcal{Z}_i,
\alpha_c,\alpha_D\right), \nonumber 
\end{eqnarray}

\noindent where ${\hat f}_e$ and ${\hat f}_i$ are universal
functions. Eqs. (\ref{DimensionlessFunctions}) already can be used to state
exact scaling laws. The requirements $\alpha_c=const$, $\alpha_D=const$
and $d/R=const$ do not fix all the dimensional parameters of the
problem, and this allows to scale experimental results. 

Yet, the most interesting scalings are obtained in the limit
$\alpha_c \ll 1$ and $\alpha_D \ll 1$. Assuming $\alpha_c\to 0$ one obtains 

\begin{equation}
\label{Equilibrium1}
\frac{{\bf \hat
p_e}}{|{\hat {\bf p_e}}|}\cdot\frac{\partial {\hat f}_e({\hat t},{\hat
{\bf p_e}}, 
{\hat {\bf r}})}{\partial {\hat {\bf r}}}-{\hat {\bf
E}}\cdot\frac{\partial {\hat f}_e({\hat t},{\hat {\bf p_e}}, 
{\hat {\bf r}})}{\partial {\hat{\bf p_e}}}=0.
\end{equation}

\noindent This means that the electron distribution function can be written as

\begin{equation}
\label{Equilibrium2}
{\hat f}_e=F_e\left(\hat{|\bf p_e|}-\hat{\phi},\hat{t},\mathcal{Z}_i,d/R,\alpha_D\right).
\end{equation}

\noindent where $F_e$ is a universal
function. Eq. (\ref{Equilibrium2}) means that the electron fluid has
the same effective temperature at all points.

The formal limit $\alpha_D\to 0$ coincides with the quasineutrality
condition 

\begin{equation}
\label{Quasineutrality} 
\mathcal{Z}_i\int{\hat f}_i\,d^3{\hat{\bf p}}=\int {\hat
f}_e\,d^3{\hat{\bf p}}). 
\end{equation}

\noindent Since $\alpha_D$ is a factor in front of the highest
derivative in (\ref{VlasovDimensionless}), the quasinetrality
condition (\ref{Quasineutrality}) is violated within the narrow Debye
sheath layer of the width $\propto \lambda_D$. Being very important
for problems like ion acceleration this area hardly plays any role in
the ion focusing. Because of its narrowness, only the small amount of
ion beam on the order of $\lambda_D/R \propto \sqrt{\alpha_D} \ll 1$
would be influenced by its fields at any particular moment. We neglect
this influence.

\section{Ion focusing by hollow plasma cylinder}

In order to describe the focusing, we study properties of the
quasineutral part of the expanding plasma cloud. The quasineutrality
$n_e\approx \mathcal{Z}_i n_i$ is guaranteed as long as $\alpha_D\to
0$. Here $n_e$ and $n_i$ are the electron and ion densities,
$\mathcal{Z}_i$ is the ion charge state. 
At the same time, the plasma density
and consequently the electron pressure $P_e = n_e T_e$ vary
along the cylinder radius. The pressure gradient is counterbalanced by
the radial electric field 

\begin{equation}
\label{Balance}
{\bf E}=-\frac{1}{e n_e}\nabla P_e,
\end{equation}

\noindent which is developed inside the plasma to  
satisfy the quasineutrality (\ref{Quasineutrality}). Because the
electron 
pressure gradient is directed off axis, the developed electric field
is directed toward the cylinder axis. It is this field that focuses the
injected ions.

Because of the cylindrical symmetry, we 
neglect any dependencies on the azimuthal angle on the longitudinal
coordinate $z$ within the plasma. Then, all distributions depend on the radius
$\rho=\sqrt{x^2 + y^2}$ only. To obtain a closed system of equations
we take into account the energy conservation law  

\begin{equation}
\label{EnergyConservation}
\frac{3}{2}T_e N_e + \pi M L \int {\bf v}^2 f_i(t,{\bf
v},{\rho})\,d{\bf v}\, \rho d{\rho}= \frac{3}{2}T_e(0) N_0, 
\end{equation}

\noindent where $N_e$ is the number of hot electrons and $(3/2) T_e(0) N_e$
is the laser energy absorbed in the cylinder and stored in the hot
electrons.  Eq. (\ref{EnergyConservation}) neglects
the energy accumulated in the electromagnetic plasma fields. This
assumption is correct provided that the Debye length is much smaller
than the cylinder radius $R$, i.e. for $\alpha_D\ll 1$.  
Eqs. (\ref{Equilibrium1}) and (\ref{Equilibrium2}) show
that the electron temperature is equal at all points of
the plasma. Thus, the energy conservation law
(\ref{EnergyConservation}) is sufficient to describe the electron
dynamics. 

The initial ion distribution is

\begin{equation}
\label{InitialCondition}
f_i(t=0,{\bf p},{\rho})=2\pi \sigma_i\delta({\bf p})\tilde{F}_0({\rho}/R,d/R).
\end{equation}

\noindent where $\sigma_i$ is the initial surface density of ions
participating in the plasma expansion. Because of the quasineutrality
condition (\ref{Quasineutrality}), we have $N_e = 2\pi RL \mathcal{Z}_i \sigma_i$.

We introduce the dimensionless time-dependent electron temperature 
$\hat{T}(\hat{t}) = T_e(t)/T_e(0)$ and the ion velocity $\hat{\bf
v}={\bf v}/c_s$.

The ion Vlasov equation and
Eqs. (\ref{Balance})--(\ref{EnergyConservation}) rewritten in the 
dimensionless variables take the form:

\begin{eqnarray}
\label{BalanceUni}
-\hat{\bf E} & = & \hat{\nabla} \ln n_e, \\
\label{IonVlasovUni}
\frac{\partial \hat{f}_i}{\partial \hat{t}}
+ \hat{{\bf v}} 
\frac{\partial \hat{f}_i}{\partial \hat{\rho}}
+ \hat{\bf E} 
\frac{\partial \hat{f}_i}{\partial \hat{\bf v}} & = & 0, \\
\label{EnergyConservationUni}
\int \hat{\bf v}^2 \hat{f}_i(\hat{t},\hat{\bf v},\hat{\rho}) \,d\hat{\bf
v}d\hat{r} & = & {3}(1-\hat{T}),
\end{eqnarray}

\noindent with the initial condition

\begin{equation}
\label{InitialConditionUni}
\hat{f}_i(t=0,\hat{\bf v},\hat{\rho})=F_0\left(\hat{\rho},d/R\right)
\delta(\hat{\bf v}).
\end{equation}

\noindent Eqs.~(\ref{BalanceUni})-(\ref{InitialConditionUni})
contain no dimensional parameters whatsoever.
As a consequence, the functions $\hat{T}$, $\hat{f}$ and $\hat{\bf E}$ are
universal, i.e., they are not affected by specific values of $d$,
$R$, $L$, $\sigma_i$ and $T_e(0)$. This gives us an opportunity to develop a
meaningful similarity theory describing the guidance of laser produced
ion beams. 

From the normalizations (\ref{DimensionlessVariables}) 
we conclude that the electric field $\bf E$ developed in the plasma is

\begin{equation}
\label{ElectricField}
{\bf E}=\frac{T_e(0)}{eR}\hat{\bf E}(t/\tau,{\rho}/R,d/R).
\end{equation}

\noindent where $\hat{\bf E}$ is a universal
function. It does not depend on the plasma density, but is
determined by the hot electron temperature and the cylinder
geometry only. This result is valid as long as the Debye length is much
smaller than $R$. This means that the 
uncompensated charge density 

\begin{equation}
\label{ChargeDensity}
e\delta n = e(\mathcal{Z}_i n_i-n_e) =\frac{\nabla {\bf E}}{4\pi}=\frac{T_e(0)}{4\pi
e R^2}\delta\hat{n}(t/\tau,{\rho}/R,d/R) 
\end{equation}

\noindent is much smaller than the electron density. 

When the laser produced ion beam enters plasma inside the cylinder, it
is deflected by the electric field
(\ref{ElectricField}). We suppose that the beam has a lower density
than the plasma inside the cylinder and thus the beam own fields can
be neglected.
%%%%%%%%%%%%%%%%%%%%%%%%%%%%%%%%%%%%%%%%%%%%%%%%%
% 

To describe the beam ion guiding in plasma we consider ions with
the charge state $\mathcal{Z}_b$, mass $M_b$ and 
the initial energy $\mathcal{E}_b$ being focused by the potential

\begin{equation}
\label{Potential}
\varphi=-\pi r^2 e \delta n_0,~~~ \delta n_0= \delta n(t/\tau,\rho=0,d/R)
\end{equation}

\noindent Notice that
the charge density $\delta n_0$ depends on time. However, 
for the most interesting and important case the beam ions pass the
cylinder plasma during the time $L/u_b \ll \tau$, where
$u_b=\sqrt{2\mathcal{E}_b/M_b}$ and $\tau=R/c_S$ is the plasma
evolution time. In this case the dependence of $\delta n$
from time $t$ can be neglected.  

Now we are able to estimate influence of the non-neutral Debye sheath
with the width $\lambda_D$ on the beam ions motion. This area
propagates with the velocity $\propto c_s$ and carries the electric field
$E_{nq} \propto\sqrt{n_eT_e}$. The radial momentum of a beam ion is
changed by the value 

\begin{equation} 
\label{MomentumChangeNQ}
\Delta p_{\perp}^{nq}\propto
eZ_i\sqrt{n_eT_e}\frac{\lambda_D}{c_s}\propto \sqrt{\Delta
\mathcal{E}_{nq} M_b}, 
\end{equation}

\noindent 
where 
$\Delta \mathcal{E}_{nq} \propto \frac{M_i}{M_b}T_e$.

\noindent 
The change of  a beam ion radial momentum due to the interaction with
the quasineutral part of plasma is estimated as

\begin{equation}
\label{MomentumChange}
\Delta p_{\perp}^q\propto \frac{T_e}{R}\frac{L}{u_b}\propto
\sqrt{\Delta \mathcal{E}_q M_b}, 
\end{equation}

\noindent
where $\Delta \mathcal{E}_q\propto
\left(\frac{L}{u_b\tau}\right)^2\frac{M_i}{M_b}T_e$.

From Eqs. (\ref{MomentumChangeNQ}) and (\ref{MomentumChange}) one
sees that $\Delta p_{\perp}^{nq}\gg \Delta p_{\perp}^q$. 
Therefore the ions passing through the non-quasineutral edge
are stronger deviated than those interacting only
with the quasineutral plasma region. However, because the Debye sheath
is narrow, the relative number of the strongly declined ions is small
and these ions are deviated to different points of space. 
For these reasons the Debye
sheath at the edge of the expanding plasma does not contribute to
the ion focusing. It scatters the beam ion instead.

\section{Ion lens formula}

To investigate focusing properties of the potential
$\varphi$ we use the well known analogy between the geometrical optics
and the classical mechanics. The optical length 
corresponds to the action $S$ in the Hamilton-Jacoby equation \cite{Arnold} 

\begin{equation}
\label{Hamiltonian}
\partial_tS+H(\nabla S,{\bf r})=0.
\end{equation}

\noindent If ions in vacuum are injected at the point $x=y=0$, $z=Z$
then the $S$ function in vacuum is 

\begin{eqnarray}
\label{Source}
S&=&-\mathcal{E}_bt+\sqrt{2M_b\mathcal{E}_b\left((z-Z)^2+\rho^2\right)}\\
&\approx&-\mathcal{E}_bt+\sqrt{2M_b\mathcal{E}_b}(z-Z)+\frac{\rho^2}{z-Z}\sqrt{\frac{M_b\mathcal{E}_b}{2}}. \nonumber
\end{eqnarray}

\noindent In our geometry, $Z=-D$. 

If ions are focused at the point $x=y=0$, $z=Z'$ in vacuum,
then the action $S$ is 

\begin{eqnarray}
\label{Source1}
S&=&const-\mathcal{E}_bt-\sqrt{2M_b\mathcal{E}_b\left((z-Z')^2+\rho^2\right)}\\
&\approx&const-\mathcal{E}_bt+\sqrt{2M_b\mathcal{E}_b}(z-Z')+\frac{\rho^2}{z-Z'}\sqrt{\frac{M_b\mathcal{E}_b}{2}}. \nonumber
\end{eqnarray}

\noindent The beam ion motion inside the plasma cylinder is described by the
Hamiltonian 

\begin{equation}
\label{HamiltonianInPlasma}
H=\frac{{\bf p}^2}{2M_b}+ e \mathcal{Z}_b \varphi.
\end{equation}

\noindent The solution of the Hamilton-Jacoby equation inside the
plasma can be expanded as 

\begin{equation}
\label{SolutionForm}
S=const-\mathcal{E}_bt+\sqrt{2\mathcal{E}_bM_b}z+\frac{1}{2}\beta(z)\rho^2+...,
\end{equation}

\noindent where the function $\beta$ is 

%%%%%%%%%%%%%%%%%%%%%%%%%%%%%%%%%%%%%%%%%%%%%%%%%%%%%%%%%%%%%%%%%%
\begin{equation}
\label{BetaFunction}
\beta(z)=\sqrt{-2\pi \mathcal{Z}_b M_b e^2 \delta n_0}\tan\left(z\sqrt{-\frac{\pi Z_b e^2 \delta n_0}{\mathcal{E}_b}}+C\right).
\end{equation} 
%%%%%%%%%%%%%%%%%%%%%%%%%%%%%%%%%%%%%%%%%%%%%%%%%%%%%%%%%%%%%%%%%%%%%%%%%%%%%%%%%%5555
\noindent Note that when the electrons pull the ions behind themselves
(positive pressure gradient) there is an electron excess at the axis
$z$, i.e., $\delta n_0<0$. 
The constant $C$ in (\ref{BetaFunction}) is obtained from the
continuity conditions of the action $S$ at the front and rear sides of
the plasma cylinder.

Thus, we  arrive at the thick lens formula  

\begin{equation}
\label{ThickLense}
(Z-g)(Z'+h)=-f^2
\end{equation}

\noindent where $g=-(L\cos\epsilon)/(\epsilon\sin\epsilon)$, $h=g+L$,
$f=L/(\epsilon\sin\epsilon)$ and 

\begin{equation}
\label{Epsilon}
\epsilon=\frac{\sqrt{\mathcal{Z}_b \delta\hat{n}(t/\tau,0,d/R)}}{2}\frac{L}{u_b\tau}
\end{equation}

\noindent %Here $u_b=\sqrt{2\mathcal{E}_b/M_b}$ is the beam ion velocity.
In our derivation of Eq. (\ref{ThickLense}) we neglected the
change of plasma parameters during the time the beam ions need to pass the
cylinder. This means that our analysis is valid if $\epsilon \ll 
1$. 

A parallel beam of ions is obtained if 
$Z'=\infty$. This
condition is satisfied for $\epsilon\ll 1$ if 

\begin{equation}
\label{ParallelBeam}
D=\frac{L}{\epsilon^2}.
\end{equation}

\noindent Thus, the plasma element collimates ions with the energy

\begin{equation}
\label{Collimation}
\mathcal{E}_b\propto \mathcal{Z}_b T_e\frac{LD}{R^2}.
\end{equation}

\noindent It is worth mentioning that the energy of the collimated ions
strongly depends on the dimensionless parameter $LD/R^2$ and can be
significantly larger than the initial electron temperature. 
%%%%%%%%%%%%%%%%%%%%%%%%%%%%%%%%%%%%%%%%%%%%%%%%%%%%%%%%%%%

It is easy to see that for the ion focusing be practical the
electron temperature $T_e$ has  
to be of the order of several MeVs. Such electron temperatures are
routinely produced by multi-terawatt lasers.  

Would be the plasma inside the cylinder stationary, then only ions
with the selected energy (\ref{Collimation}) are collimated. However,
the plasma is non-stationary with the characteristic evolution time
$\tau=R/c_s$. The relative change of the plasma parameters during the
ion passage time through the cylinder is of the order of
$L/v_b\tau \ll 1$. This small parameter defines finally the
finite energy spectrum width $\delta \mathcal{E}_b$ of the focused
ions:

\begin{equation}
\label{WidthEnergy}
\frac{\delta \mathcal{E}_b}{\mathcal{E}_b}\propto \frac{L}{v_b\tau}\propto \sqrt{\frac{L}{D}}.
\end{equation}

\noindent It follows from (\ref{WidthEnergy}) that the plasma cylinder
works as a good monochromator if $D\gg L$.

To avoid any confusion we emphasize that Eq. (\ref{WidthEnergy})
describes the quality of a small aperture ion beam only. Of course, 
different parts of the lens collimate ions of different
energies. For large aperture ion beams the energy spectrum width of
the focused ions will be large,  $\Delta \mathcal{E}_b/\mathcal{E}_b
\propto 1$.

In the preceding part of the paper we consider the focusing by the
area $\rho\approx 0$. It is important that the theory can easily be
generalised for the focusing by the cylindric area near
$\rho=\rho_0<R$. To do so we introduce a new potential  

\begin{equation}
\label{PotentialR_0}
\varphi_{\rho_0}=-\pi r^2 e \delta n_{r_0},~~~ \delta n_{\rho_0}=\frac{T_e(0)}{2\pi e^2\rho_0R}\hat E(t/\tau,{\rho}_0/R,d/R). 
\end{equation}

\noindent According to Eq. (\ref{ElectricField}) the potential
$\varphi_{r_0}$ gives the right value of the electric field at
$\rho=\rho_0$. Thus the focusing by $\rho=\rho_0$ is obtained from
Eq. (\ref{ThickLense}) with $\epsilon\ll1$ by the substitution $\delta
n_0\to \delta n_{r_0}$.  

Until now we have assumed that the density of the ion beam focused is
so small that it does not affect the focusing field of the lens. To
find the validity condition for this approach we have to consider the
propagating of an ion beam with a given density profile $n_b(\rho)$
through the lens. Using the quasuneutrality condition for the system
"the lens plasma + the beam" one can easily find the the focusing by
the area around $\rho=\rho_0$ is not disturbed by the ion beam if 

\begin{equation}
\label{Validity}
|\partial_{\rho} n_b(\rho_0)|\ll|\partial_{\rho} n_i(t/\tau,\rho_0)|.
\end{equation}

\noindent Since the plasma lens density gradient can be very large this condition can be much weaker than $n_b\ll n_i$.

\section{Conclusions}

In conclusion, we have developed a closed similarity theory of a
hollow cylinder as a plasma element for ion beam
guiding. Significantly, the beam ions are focused by the quasinuetral part
of expanding plasma rather than by strong electric fields in 
the non-quasineutral leading edge of the expanding plasma cloud.
The thick lens formula has been obtained with explicit scalings for all of the
parameters. We show that the plasma lens collimates only ions with a
quite definite energy and may be used for monochromatization of the
laser-produced ion beams.

\section*{Aknowledgements}

This work has been supported in parts by Transregio-18 and
Graduiertenkolleg 1203 (DFG, Germany).

\end{document}